\documentclass[11pt]{article}
\usepackage[utf8x]{inputenc}
\usepackage{amssymb,amsmath,mathrsfs,enumerate}
\usepackage{graphicx,rotate,multicol}
\usepackage{float}
\usepackage{tocloft}
\usepackage{subfig}
\usepackage{multirow}
\usepackage{slashed}
\usepackage[margin=10pt,labelfont=bf]{caption}
\usepackage{cite}
\usepackage{soul}
\usepackage[colorlinks=true,
linkcolor=blue,
urlcolor=blue,
citecolor=teal]{hyperref}

\makeatletter
\let\Hy@linktoc\Hy@linktoc@page
\makeatother

\usepackage{color}
\definecolor{ourcolor}{rgb}{0.7, 0.25, 0.05}
\usepackage{tikz,braket}
\long\def\rpl#1!!#2!!{\textcolor{red}{#1} \textcolor{blue}{#2}}

\let\tilde=\widetilde

\let\bar=\overline

\def \order(#1){{\mathcal O} \left(#1 \right)}

\evensidemargin=\oddsidemargin
\allowdisplaybreaks

\title{\color{black}{\bf Status of a Flavour Maximal Non-minimal Universal Extra Dimension}}

\author {\bf Sayan Dasgupta,$^{a,}$\footnote{sayandg05@gmail.com}
	\hspace{4pt} Ujjal Kumar Dey,$^{a,}$\footnote{ujjal@cts.iitkgp.ernet.in} 
	\hspace{4pt}  Tapoja Jha$^{b,c,}$\footnote{tapoja.j@iopb.res.in}
	\hspace{4pt} Tirtha Sankar Ray$^{a,d,}$\footnote{tirthasankar.ray@gmail.com} \\[10pt]
	\small\em $^a$Centre for Theoretical Studies, 
	Indian Institute of Technology Kharagpur,
	Kharagpur 721302, India\\
	\small\em $^b$Institute of Physics, 
	Sachivalaya Marg, 
	Bhubaneswar, Odisha 751005, India\\
	\small\em $^c$Homi Bhabha National Institute,
	Training School Complex, Anushakti Nagar, 
	Mumbai 400085, India\\
	\small\em $^d$Department of Physics, Indian Institute of Technology, Kharagpur 721302, India              
}

\date{}

\begin{document}
	
	\maketitle
	
	\begin{abstract}
		In this paper we consider an $S^{1}/\mathbb{Z}_2$  compactified flat extra dimensional scenario where all the standard model states can access the bulk and  have generalised brane localised kinetic terms.  The flavour structure of  brane kinetic terms  for the standard model fermions are dictated by stringent flavour bounds on  the first two  generations  implying an $U(2)_{Q_L} \otimes U(2)_{u_R} \otimes U(2)_{d_R}$  flavour symmetry. We consider the  constraints  on  such a scenario arising from   dark matter  relic density and direct detection measurements,  precision electroweak data, Higgs physics and LHC dilepton searches. We discuss the possibility of such a scenario providing an explanation   of the recently measured anomaly in $R_{K^{(\ast)}}$ within the allowed region of the parameter space.
	\end{abstract}
	
	
	
	
	\newpage
	
	\hrule \hrule
	\tableofcontents
	\vskip 10pt
	\hrule \hrule 
\section{Introduction}
\label{sec:intro}

Universal Extra Dimension (UED) \cite{Antoniadis:1990ew, Antoniadis:1998ig, Appelquist:2000nn} as an extension of the Standard Model (SM) of particle physics has received considerable attention in the literature, see \cite{Hooper:2007qk} and references therein. In these models, one additional spatial dimension  with a flat metric is considered compactified on an $S^1/\mathbb{Z}_2$ orbifold. The end points where translation symmetry is explicitly broken are locations of two four dimensional space-time hyper-surfaces called the 3-branes. The reflection symmetry of the bulk geometry results in a conserved Kaluza-Klein (KK) parity that can stabilize a Dark Matter (DM) candidate in this setup providing the strongest motivation. In the minimal version of this model (mUED) upper bound on the observed relic density places the scale of new physics at  the TeV  scale  \cite{Servant:2002aq, Kong:2005hn} implying encouraging prognosis of being explored at collider experiments. Interestingly, the collider phenomenology~\cite{Dey:2014ana, Choudhury:2016tff, Beuria:2017jez, Chakraborty:2017kjq} of these models closely mimics supersymmetric extension of the SM with a relatively compressed spectrum \cite{Cheng:2002ab}. 

The radiative corrections  modify  both the masses and couplings partially lifting the degeneracy in the model \cite{Cheng:2002iz}. The non-renormalizability of the 5d theory ensures that the radiative corrections are proportional to the cutoff and thus incalculable. However a prudent way to accommodate such corrections is to introduce brane localized kinetic terms (BLKT)
\cite{Flacke:2008ne, delAguila:2003bh, delAguila:2003kd, delAguila:2003gu, delAguila:2003gv, delAguila:2006atw, Flacke:2014jwa}. The BLKT parameters are eventually related to the radiative corrections in a UV complete model, however in this paper we will consider them to be free parameters in the spirit of the so  called non-minimal  UED models (nmUED). Myriad phenomenological aspects of such a setup has been studied including LHC searches of the strong sector~\cite{Datta:2012tv, Datta:2013yaa}, Higgs data~\cite{Dey:2013cqa}, flavour physics~\cite{Datta:2015aka, Datta:2016flx, Biswas:2017vhc}, unitarity bounds~\cite{Jha:2016sre}, $Z\to b\bar{b}$ decay width~\cite{Jha:2014faa}, rare top decays~\cite{Dey:2016cve} and some other sectors~\cite{Datta:2012xy, Flacke:2013pla, Datta:2013lja, Shaw:2014gba, Shaw:2017whr}. If  identical BLKT parameters are introduced in the two branes for every bulk field, the KK symmetry  is preserved and can lead to a stable DM  with a  rich phenomenology \cite{Datta:2013nua, Flacke:2017xsv}.

The BLKT parameters having its origin in the  radiative corrections are expected to be dependent on flavour owing to the  mass hierarchy in the SM fermions. However, most studies in the literature have remained confined to universal BLKT parameters owing to the dangerous tree level FCNC and level mixing that can arise in the case of most generalized BLKT parameters. Taking cue from some recent hints about flavour violation beyond the SM in $B$ meson decays \cite{Aaij:2014ora},  in this paper we explore a possible extension of this setup  while imposing an  $U(2)_{Q_L} \otimes U(2)_{u_R} \otimes U(2)_{d_R}$ flavour symmetry on the BLKT parameters. 
This ensures the absence of  tree level FCNC mediated by SM gauge bosons in the first two generations of the quark sector. Additionally, in the limit where neutrino masses are neglected, the absence of tree level FCNC is ensured in the leptonic sector also~\cite{Megias:2017vdg}.  We also keep the BLKT for bosonic degrees of freedom to be identical to prevent level mixing in the gauge sector that can lead to nontrivial constraints form the oblique electroweak parameters. Finally we assume a universal brane term for the Yukawas that suppresses flavour violation in the scalar sector while leaving enough freedom in the 5d theory to reproduce the CKM matrix. This setup represents the  \textit{flavour maximal nmUED}.

The non-observation of any hint of new physics at the Large Hadron Collider (LHC) either in direct searches for resonances \cite{atlasExotic, cmsExotic} or in the increasingly SM like Higgs couplings \cite{Khachatryan:2016vau} has been pushing the scale of new physics ever  higher. This leads to tension with the upper limit on the scale of the extra dimension coming from the over closure bound for dark matter  relic density. In this paper we explore the possibility  of the allowed parameter space in addressing the observed   discrepancy in the recent  measurement of $R_{K^{(\ast)}} (= \mbox{Br}(\bar{B}\to \bar{K^{(\ast)}}\mu \mu)/ \mbox{Br}(\bar{B}\to \bar{K^{(\ast)}}e e))$ at LHCb~\cite{Aaij:2014ora}. We perform an extensive scan of the  parameter space to find the regions of parameter space that is in consonance with the constraints from flavour \cite{Aaij:2014ora, Lees:2012xj, Huschle:2015rga, Aaij:2015yra}, dark matter relic density and direct searches \cite{Akerib:2017kat, Aprile:2018dbl}, Higgs data \cite{Ellis:2013lra}, precision electroweak parameters \cite{Baak:2014ora} and LHC constraints from the  dilepton channel \cite{ATLAS-CONF-2016-045, CMS-PAS-EXO-16-031}. We find only a very tuned region of parameter space with some large BLKT parameters survive the onslaught.

The rest of the paper is arranged as follows. In the next section we briefly sketch the model and set up the parameter space. In Sec.~\ref{sec:RDRK} we discuss the constraints from the  flavour observables $R_{K^{(\ast)}}$ and $R_{D^{(\ast)}}.$  Dark matter relic density and direct detection observations are discussed in Sec.~\ref{sec:dm}. In Sec.~\ref{sec:phenoconstraints} the phenomenological constraints from Higgs data, oblique parameters, LHC dilepton are briefly reviewed. We then present the results of our extensive parameter scan and discuss their phenomenological implications  in Sec.~\ref{sec:results} before concluding  in Sec.~\ref{sec:concl}. We give a few finer details of the model and the relevant flavour violation in the appendix.


\section{The Maximally Flavoured nmUED}
\label{sec:model}
We will consider that the extra spatial dimension is compactified on an  $S^{1}/\mathbb{Z}_2$   and all standard model states can acess the bulk. The end points are locations of the 3-branes with symmetric brane localized terms.
Schematically the  five-dimensional action for the SM quarks is given by,
\begin{eqnarray}
\label{eq:quarkAction}
\mathcal{S}_{{\rm quark}} &=& \int d^4 x \int_{0}^{\pi R} dy \Big[i\overline{Q}_{k} \Gamma^{M} \mathcal{D}_{M} Q_{k} + r_{f_{k}} \{ \delta(y) + \delta(y-\pi R) \} i\overline{Q}_{k} \gamma^{\mu} \mathcal{D}_{\mu} P_L Q_{k} \nonumber \\ 
 & & + i\overline{U}_{k} \Gamma^{M} \mathcal{D}_{M} U_{k} + r_{f_{k}} \{ \delta(y) + \delta(y-\pi R) \} i\overline{U}_{k} \gamma^{\mu} \mathcal{D}_{\mu} P_R U_{k} \nonumber \\ 
 & & + i\overline{D}_{k} \Gamma^{M} \mathcal{D}_{M} D_{k} + r_{f_{k}} \{ \delta(y) + \delta(y-\pi R) \} i\overline{D}_{k} \gamma^{\mu} \mathcal{D}_{\mu} P_R D_{k}\Big], 
\end{eqnarray}
where the subscript $k$ represents the flavour index. We follow similar prescription for leptons. Further details of the model including the KK decomposition leading to the KK towers for all bulk fields, is given in appendix~\ref{appn:model}.
The masses of the KK excitation for SM quarks, represented by $M_{Qn},$ can be obtained by solving the following transcendental equations,
\begin{eqnarray} \label{eq:transc}
  r_{f} M_{Qn}= \left\{ \begin{array}{rl}
         -2\tan \left(\frac{M_{Qn}\pi R}{2}\right) &\mbox{$\forall$ $n$ even,}\\
          2\cot \left(\frac{M_{Qn}\pi R}{2}\right) &\mbox{$\forall$ $n$ odd.}
          \end{array} \right.         
 \end{eqnarray}
Evidently, for $r_{f} = 0$ this reduces to the UED KK-mass $n/R$.
Following the notation of \cite{Dey:2016cve}, the  gauge, scalar and Yukawa sector Lagrangians are given by,
\begin{align}
\label{eq:gaugeacn}
\mathcal{S}_{\rm gauge} &= -\frac14 \int d^{4}x \int_{0}^{\pi R}dy
             \bigg[
                  \sum_{a}\left(\mathcal{F}^{MNa}\mathcal{F}_{MN}^{a}
                  +r_{g}\{\delta(y)+\delta(y-\pi R)\}
                  \mathcal{F}^{\mu \nu a}\mathcal{F}_{\mu \nu}^{a}\right) \nonumber
                   \\
             & ~~~~~~~~~~~~~~~~~~~~~~~~~~~~~~ 
                   +\mathcal{B}^{MN}\mathcal{B}_{MN}+r_{g}
                   \{\delta(y)+\delta(y-\pi R)\}
                   \mathcal{B}^{\mu \nu}\mathcal{B}_{\mu \nu}
              \bigg],\\
\label{eq:scalacn}
\mathcal{S}_{\rm scalar} &= \int d^4 x \int_{0}^{\pi R} dy \Big[\left(\mathcal{D}^{M}\Phi\right)^{\dagger}\left(\mathcal{D}_{M}\Phi\right) +\tilde{\mu}_{h}^{2}\Phi^{\dagger}{\Phi}-\tilde{\lambda}_{h}{(\Phi^{\dagger}{\Phi})}^{2} \nonumber \\
 & ~~~~~~~~~~~~~~~~~~~~~~~~~~~~~~
                   + \{ \delta(y) + \delta(y - \pi R)\}\left(r_{\phi}\left(\mathcal{D}^{\mu}\Phi\right)^{\dagger}\left(\mathcal{D}_{\mu}\Phi\right) \right. \nonumber \\
& ~~~~~~~~~~~~~~~~~~~~~~~~~~~~~~
                    + \left. \mu_{B}^{2}\Phi^{\dagger}{\Phi}-\lambda_{B}{(\Phi^{\dagger}{\Phi})}^{2} \right) \Big],\\
\label{eq:yukacn}               
\mathcal{S}_{\rm Yuk} &= - \int d^{4}x \int_{0}^{\pi R}dy 
             \bigg[
             \tilde{y}^{u}_{ij}\bar{Q}_{i}\tilde{\Phi}U_{j} +
             \tilde{y}^{d}_{ij}\bar{Q}_{i}\Phi D_{j} + 
             r_{y}\{\delta(y)+\delta(y-\pi R)\} \nonumber \\
       &  ~~~~~~~~~~~~~~~~~~~~~~~~~~~~~~ 
             \times \left(
             \tilde{y}^{u}_{ij}\bar{Q}_{iL}\tilde{\Phi}U_{jR} +
             \tilde{y}^{d}_{ij}\bar{Q}_{iL}\Phi D_{jR} \right) + {\rm h.c.}
             \bigg].              
\end{align}
In this study we will assume universal BLKT parameter for electroweak gauge and scalar sectors, i.e., $r_{g} = r_{\phi}$ which simplifies the gauge sector~\cite{Muck:2004zz, Datta:2013yaa, Datta:2014sha}. Moreover, in the fermionic sector we will consider flavour dependent BLKT parameters modulo  a global flavour symmetry of $U(2)_{Q_L} \otimes U(2)_{u_R} \otimes U(2)_{d_R}.$  This suppresses dangerous flavour changing neutral (FCNC) and charged currents (FCCC) in the quark sector which is highly constrained from various flavour observations e.g., $K-\bar{K}$ oscillation, $\epsilon$-parameter etc.~\cite{Buras:1996cw}. The Yukawa couplings explicitly break this symmetry leading to an acceptable minimal flavour violation framework~\cite{Isidori:2012ts}.  We consider  $r_{g} > r_{f_{i}}$ to ensure that the lightest KK particle (LKP) is the  $B^{(1)}_\mu.$ The BLKT set-up used in the phenomenological studies in this paper has been summarised in the Table~\ref{tab:blt}. We will present all our results in terms of the  scaled dimensionless  BLKT variables $R_{X}$ where $R_{X} = r_{X}/R$. For $R_{X} < -\pi$  the masses obtained from Eq.~\eqref{eq:transc} becomes imaginary  giving  rise to tachyonic zero modes~\cite{Datta:2012tv, Datta:2013yaa}. Throughout the paper we  will restrict ourselves to ranges for the BLKT parameters from 0 to 8. The choice of the BLKT parameters, are explicitly presented in Table~\ref{tab:blt}.
\begin{table}[t]
\centering
\begin{tabular}{|c|c||c|c|}
\hline
$\begin{pmatrix}\nu_{e}\\ e\end{pmatrix}_{L}$, $e_{R}$             & $r_{f_{1}}$ & $\begin{pmatrix}u\\ d \end{pmatrix}_{L}$, $u_{R},~d_{R}$ & $r_{f_{1}}$ \\ \hline
$\begin{pmatrix}\nu_{\mu}\\ \mu \end{pmatrix}_{L}$, $\mu_{R}$      & $r_{f_{2}}$ & $\begin{pmatrix}c\\ s \end{pmatrix}_{L}$, $c_{R},~s_{R}$ & $r_{f_{1}}$ \\ \hline
$\begin{pmatrix}\nu_{\tau}\\ \tau \end{pmatrix}_{L}$, $\tau_{R}$ & $r_{f_{3}}$ & $\begin{pmatrix}t\\ b \end{pmatrix}_{L}$, $t_{R},~b_{R}$ & $r_{f_{3}}$ \\ \hline
$W_{\mu}$, $B_{\mu}$, $G_{\mu}$, $H$                               & $r_{g}$     & Yukawa                                                   & $r_{y}$     \\ \hline
\end{tabular}
\caption{Choice of BLT parameters.}
\label{tab:blt}
\end{table}
In the KK parity conserving scenario the even KK mode gauge bosons have couplings with the SM fermions, i.e., the zero mode fermions. In the gauge basis the effective Lagrangian can be written as,
\begin{align}
\label{eq:LEW}
\mathcal{L}_{\rm X} = \sum_{X = \gamma, Z} \frac{X_{\mu}^{(n)}}
                       {2c_{w}}\left[
                       g_{f_{L}}^{X^{(n)}}\bar{f}_{iL}\gamma^{\mu}f_{iL}
                       + g_{f_{R}}^{X^{(n)}}\bar{f}_{iR}\gamma^{\mu}f_{iR}
                       \right],
\end{align}
where $c_{w}$ is the usual cosine of the Weinberg angle, and the couplings are explicitly given as,
\begin{subequations}
\label{eq:Zycouplings}
\begin{gather}
g_{f_{L}}^{Z^{(n)}} = 2(T_{3f}-Q_{f}gs_{w}^{2})
                       I^{(n)},
\quad 
g_{f_{R}}^{Z^{(n)}} = -2Q_{f}gs_{w}^{2} I^{(n)},\\
g_{f_{L,R}}^{\gamma^{(n)}} = 2Q_{f}gs_{w}c_{w}
                       I^{(n)},
\end{gather}
\end{subequations}
Note that since we are not taking different BLKT parameters for left- and right-handed fermions $\gamma^{(n)}$ will have the same coupling with both. The overlap integral $I^{(n)}$ is given by,
\begin{align}
I^{(n)} = \frac{\sqrt{R_{g} + \pi}}{R_{f_{i}} +\pi}
          \sqrt{\frac{2}{\pi}}
          \frac{R_{f_{i}} - R_{g}}
               {\sqrt{1 + \frac{R_{g}^{2}
                 M_{\Phi n}^{2}}{4} 
                 + \frac{R_{g}}{\pi }}}
\label{eq:ovint}
\end{align}
Evidently, in Eqs.~(\ref{eq:LEW}) and (\ref{eq:Zycouplings}), the KK mode $n$ is even due to KK parity conservation. 
The interaction between KK modes of $W$-boson and SM fermions are given by,
\begin{align}
\label{eq:LW}
\mathcal{L}_{\rm W} = \frac{1}{\sqrt{2}} W_{\mu}^{(n)}\left[
                       g^{\prime W}_{d_{L}^{j}}\bar{u}_{j}P_{L}\gamma^{\mu}d_{j} +
                       g^{\prime W}_{\ell_{L}^{j}}\bar{\ell}_{j}P_{L}\gamma^{\mu}\nu_{j}
                       \right], 
\end{align}
where $g^{\prime W}_{d_{L}^{j}}$ and $g^{\prime W}_{\ell_{L}^{j}}$ are given by $g I(R_{g},R_{f_{j}},n)$, which are actually the overlap integral times the gauge couplings. Note that we take the BLKT parameters for the same generation of quarks and leptons to be the same. 
The Lagrangians in Eqs.~(\ref{eq:LEW}) and (\ref{eq:LW}) are in the gauge basis. In the mass basis this leads to FCNC and FCCC, mediated by the KK gauge bosons. This is because of the matrix 
\begin{align*}
G^{(n)} = \text{diag}\left(g_{f_{1}}^{X^{(n)}/W^{(n)}}, g_{f_{2}}^{X^{(n)}/W^{(n)}}, g_{f_{3}}^{X^{(n)}/W^{(n)}}\right),
\end{align*}
being diagonal but not proportional to the identity matrix. However, note that the SM GIM mechanism would still be in effect because $G^{(0)}$ is still proportional to the identity matrix owing to the orthonormality of the mode functions. As far as the leptonic sector is concerned, KK gauge boson mediated FCNC is proportional to the neutrino Yukawa couplings. In this paper we will neglect the neutrino masses and thus avoiding the FCNC in the leptonic sector. Owing to the different BLKT parameters for the lepton generations the strength of the KK gauge boson couplings to the SM charged leptons would be dependent on the generational index. In the appendix~\ref{appn:flav} we give the details of the flavour violation in the gauge and Yukawa sector.

\section{Flavour Observables: $R_{K^{(\ast)}}$ and $R_{D^{(\ast)}}$}
\label{sec:RDRK}
\begin{figure}[t]
	\centering
	\subfloat[\label{sfa:RKvsmKK}]{
		\includegraphics[scale=0.5]{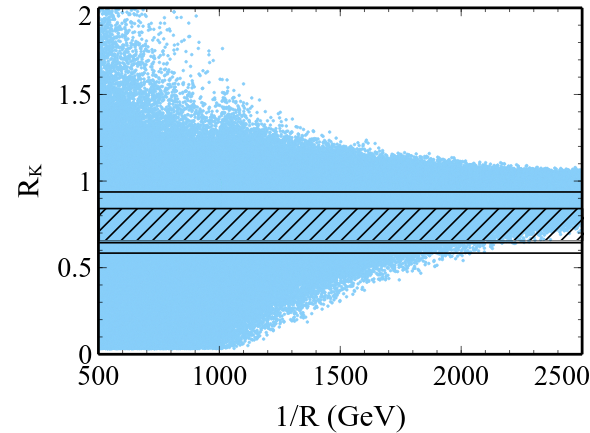}
	} ~~~
	\subfloat[\label{sfb:RKvsmKK}]{
		\includegraphics[scale=0.5]{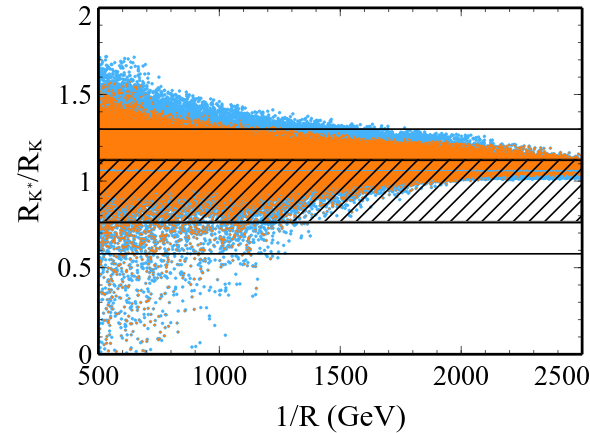}
	}
	\caption{Allowed parameter points in the (a) $R_K$-$1/R$ and (b) $R_{K^{\star}}/R_K$-$1/R$ plane. The black hatched region corresponds to 1$\sigma$ deviation from experimental results and the region enclosed by the uppermost and lowermost black lines corresponds to 2$\sigma$ deviation from experimental results. In (b), blue points denote those allowed by 2$\sigma$ of $R_K$ and orange points denote those allowed by 1$\sigma$ of $R_K.$}
	\label{RKvsmKK}
\end{figure}

Recent measurement of the flavour observables $R_{K^{(\ast)}}$ and $R_{D^{(\ast)}}$ shows a consistent $\sim 2\sigma$ discrepancy with the SM. This has stimulated a plethora of models both from bottom-up~\cite{Calibbi:2015kma, Bardhan:2016uhr, Bhattacharya:2016zcw, Ghosh:2017ber, Bardhan:2017xcc, Capdevila:2017bsm, Choudhury:2017qyt, Choudhury:2017ijp, Jaiswal:2017rve, Bhattacharya:2018kig} and top-down approach~\cite{Crivellin:2015era, Das:2016vkr, Crivellin:2017zlb, Hiller:2017bzc, Calibbi:2017qbu, DAmbrosio:2017wis, Blanke:2018sro} to explain this tension. In this section we briefly discuss the impact of these observables as well as $R_{D^{(\ast)}}$ measurements on the parameter space of the model. 
The current LHCb measurements of $R_{K^{(\ast)}}$ are given by~\cite{Aaij:2014ora, Aaij:2017vbb},
\begin{subequations}
\begin{gather}
R_{K^{(\ast)}} = \frac{\mbox{Br}(\bar{B}\to 
                 \bar{K}^{(\ast)}\mu \mu)}
                 {\mbox{Br}(\bar{B}\to \bar{K}^{(\ast)}e e)}, \\
R_{K} = 0.745^{+0.090}_{-0.074} \pm 0.036~, \quad
R_{K^{\ast}} = 0.69^{+0.11}_{-0.07} \pm 0.05~.      
\end{gather}
\end{subequations}
Integrating out the heavy KK modes from the interaction terms presented in Eq.~(\ref{eq:LEW}) we can obtain the relevant four fermion operators which contribute in the flavour observable $R_{K^{(\ast)}}$ which is basically a $\Delta F = 1$ transition. Following the standard definitions of Wilson coefficients $C_{9,10}^{(')}$~\cite{Buchalla:1995vs} we can write the new physics contributions ($\Delta C_{9,10}^{(')}$) in our model as,
\begin{subequations}
\begin{align}
\Delta C_{9} = -\sum_{\substack{X = \gamma,Z \\ \{n\}}}
                \frac{c g_{\mu_{V}}^{X^{(n)}}}{M_{X^{(n)}}^{2}}
                \left(g_{b_{L}}^{X^{(n)}}  
                 - g_{s_{L}}^{X^{(n)}}\right),
~~ 
\Delta C_{9}^{'} = -\sum_{\substack{X = \gamma,Z \\ \{n\}}}
                \frac{c g_{\mu_{V}}^{X^{(n)}}}{M_{X^{(n)}}^{2}}
                \left(g_{b_{R}}^{X^{(n)}}  
                 - g_{s_{R}}^{X^{(n)}}\right),\\
\Delta C_{10} = \sum_{\substack{X = \gamma,Z \\ \{n\}}}
                \frac{c g_{\mu_{A}}^{X^{(n)}}}{M_{X^{(n)}}^{2}}
                \left(g_{b_{L}}^{X^{(n)}}  
                 - g_{s_{L}}^{X^{(n)}}\right),
~~ 
\Delta C_{10}^{'} = \sum_{\substack{X = \gamma,Z \\ \{n\}}}
                \frac{c g_{\mu_{A}}^{X^{(n)}}}{M_{X^{(n)}}^{2}}
                \left(g_{b_{R}}^{X^{(n)}}  
                 - g_{s_{R}}^{X^{(n)}}\right),
\end{align}
\label{eq:wilsoncoeff}
\end{subequations}
where the common factor $c = \pi/(2\sqrt{2}G_{F}\alpha c_{w}^{2})$, and $g_{f_{V,A}}^{X^{(n)}} = \left(g_{f_{L}}^{X^{(n)}} \pm g_{f_{R}}^{X^{(n)}}\right)/2$. In these equations the sum over even KK modes $n$ is also taken. Now, in terms of these Wilson coefficients $R_{K}$ is given by~\cite{Megias:2017ove},
\begin{align}\label{rk}
R_{K} = \frac{\sum_{j=9,10}\left|C_{j}^{\rm SM} 
              + \Delta C_{j}^{\mu} 
              + \Delta C_{j}^{'\mu}\right|^{2}}
             {\sum_{j=9,10}\left|C_{j}^{\rm SM} 
              + \Delta C_{j}^{e} 
              + \Delta C_{j}^{'e}\right|^{2}},
\end{align} 
where $C_{9}^{\rm SM} \simeq -C_{10}^{\rm SM} \simeq 4.2$~\cite{Megias:2017vdg}. Also, following~\cite{Hiller:2014ula, Hiller:2017bzc} $R_{K^{\ast}}$ can be presented as,

\begin{align}\label{rkstarbyrk}
\frac{R_{K^{\ast}}}{R_{K}} = 1 + 
             \frac{1}{(C^{\rm SM})^{2}R_{K}}\left[
              2C^{\rm SM} 
              \left\lbrace
               \mathscr{A}^{\mu} - (\mu \to e)                
              \right\rbrace
            + \left\lbrace 
               \left|\mathscr{A}^{\mu}\right|^{2}
              + \left|\mathscr{B}^{\mu}\right|^{2}
              - (\mu \to e)
              \right\rbrace 
              \right] \;,
\end{align}
where $C^{\rm SM} = C_{9}^{\rm SM} + C_{10}^{\rm SM}$, and
%
$\mathscr{A}^{\mu} = \Delta C_{9}^{\mu} 
                     + \Delta C_{9}^{\prime \mu}
                     - \Delta C_{10}^{\prime \mu}
                     - \Delta C_{10}^{\prime \mu}\;\; ;
\mathscr{B}^{\mu} = \Delta C_{9}^{\mu} 
                     + \Delta C_{9}^{\prime \mu}
                     + \Delta C_{10}^{\prime \mu}
                     + \Delta C_{10}^{\prime \mu}\;.$
%
Within the experimental uncertainties, this ratio can be evaluated to be $R_{K^{\ast}}/R_{K} = 0.94\pm 0.18$~\cite{Hiller:2017bzc}.
\begin{figure}[t]
\centering
\subfloat[\label{sfa:b2smu}]{
\includegraphics[scale=0.5]{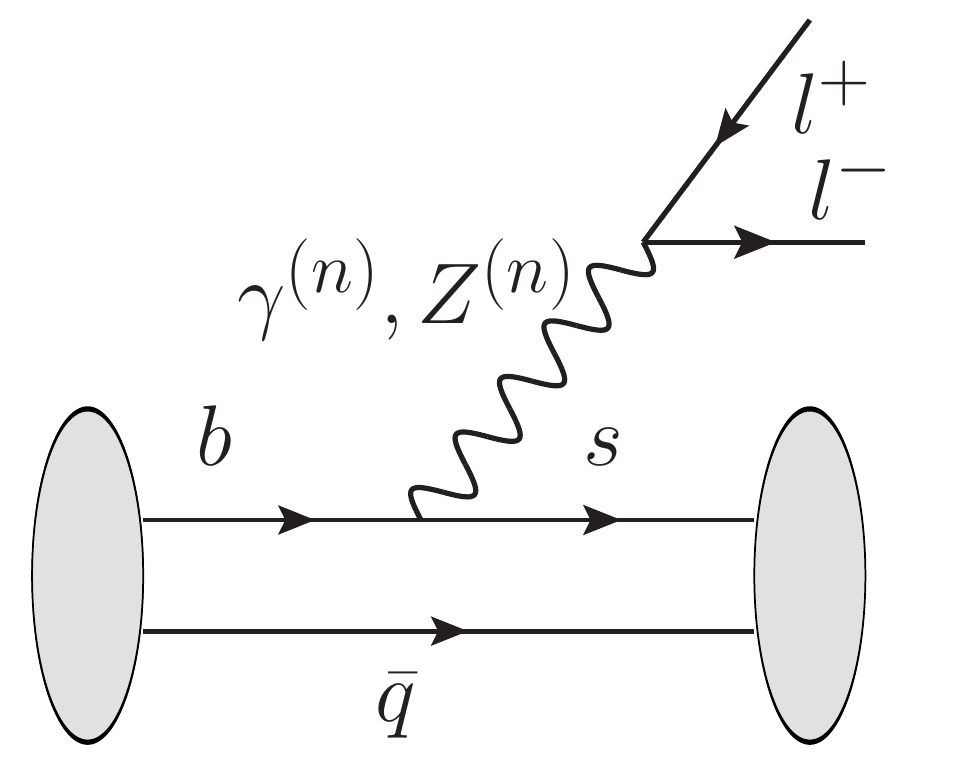}}
\subfloat[\label{sfb:b2smu}]{
\includegraphics[scale=0.5]{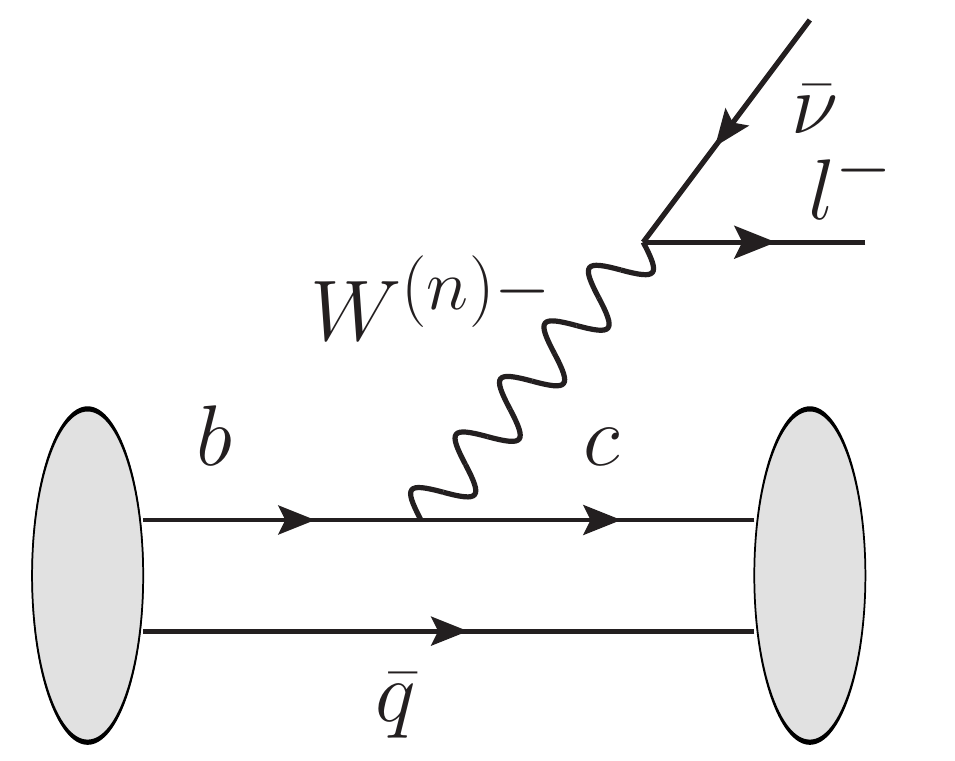}}
\caption{Schematic Feynman diagram for the process (a) $b\to s\mu\bar{\mu}~(e\bar{e})$ and (b) $b\to c l\bar{\nu}_{l}$.}
\label{fig:b2smu}
\end{figure}
We generate BLKT parameters randomly in the range $(0, 8)$ and $1/R$ in the range $(500, 2500)$ GeV and calculate $R_K$ using Eq.~\eqref{rk} and $R_{K^{\ast}}/R_K$ using Eq.~\eqref{rkstarbyrk} for each set. The resulting values have been plotted in Figure~\ref{RKvsmKK}. Within the left-right symmetric  BLKT framework (i.e., the BLKT parameters for left- and right-handed fermions are identical) considered here the primed Wilson coefficients (see Eq.~\eqref{eq:wilsoncoeff}) are not vanishing. However, we obtain large number of parameter points that simultaneously satisfy $1\sigma$ ($2\sigma$) bounds from $R_K$ and $R_{K^{\ast}}.$
We now consider the flavour observables $R_{D^{(\ast)}}$ in this model. The standard definition of which is
\begin{equation}
R_{D^{(\ast)}} = \frac{\mbox{Br}(\bar{B}\to D^{(\ast)}\tau \nu_{\tau})}
        {\mbox{Br}(\bar{B}\to D^{(\ast)}\ell \nu_{\ell})},~~
        (\ell = e ~\mbox{or}~ \mu)~. 
\end{equation}
The world average of $R_{D^{(*)}}$ from experimental results from BaBar, Belle and LHCb~\cite{Lees:2012xj, Huschle:2015rga, Aaij:2015yra} is given by~\cite{RDAvgFPCP:2017},
\begin{align}
R_{D} = 0.407\pm 0.039\pm 0.024,\quad 
R_{D^{\ast}} = 0.304\pm 0.013 \pm 0.007~.
\end{align} 
The contribution to $R_{D^{(\ast)}}$ comes after integrating out the KK modes of $W$-boson. The effective operator can be written as,
\begin{align}
\label{eq:LWeff}
\mathcal{L}_{W}^{\rm eff} = -\frac{4G_{F}}{\sqrt{2}} V_{cb}
                            \sum_{\substack{\ell \\ \{n\}}}
                            C_{\ell}^{(n)}
                            (\bar{c}\gamma_{\mu}P_{L}b)
                            (\bar{\ell}\gamma^{\mu}\nu_{\ell})\;,
\end{align}
where the KK index $n$ is even and the Wilson coefficients $C_{\ell}^{(n)}$ are given by~\cite{Megias:2017vdg},
\begin{align}
\label{eq:Cell}
C_{\ell}^{(n)} = \left(\frac{m_{W}}{m_{W^{(n)}}}\right)^{2}
                 g_{\mu_{L}}^{\prime} g_{c_{L}}^{\prime}\; .
\end{align}
In terms of the Wilson coefficients defined in Eqs.~\eqref{eq:LWeff} and \eqref{eq:Cell}, $R_{D^{(\ast)}}$ can be written as~\cite{Bhattacharya:2016mcc, Megias:2017vdg},
\begin{align}
R_{D^{(\ast)}}(C_{\tau},C_{\mu}) = 2 R_{D^{(\ast)}}^{\rm SM}
                      \frac{|1 + C_{\tau}|^{2}}
                           {1 + |1 + C_{\mu}|^{2}}\; ,
\end{align}
where the SM values are given by~\cite{Fajfer:2012vx, Bigi:2016mdz, Aoki:2016frl},
\begin{align*}
R_{D}^{\rm SM} = 0.300 \pm 0.008, \quad
R_{D^{\ast}}^{\rm SM} = 0.252 \pm 0.003\;.
\end{align*}
\begin{figure}
\centering
 \subfloat[]{
  \includegraphics[scale=0.5]{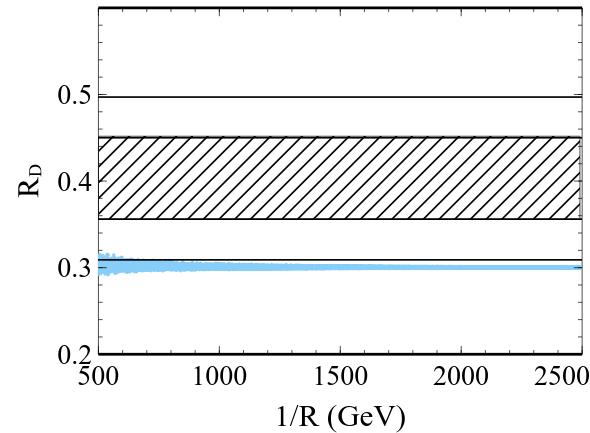}
 } ~~~
 \subfloat[]{
  \includegraphics[scale=0.5]{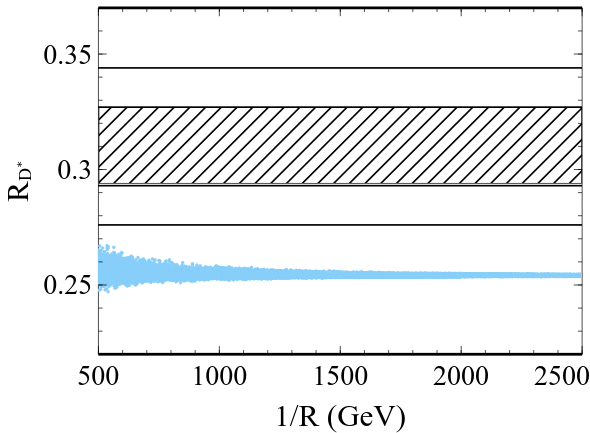}
 }
\caption{Allowed parameter points from flavour observables (a) $R_{D}$ and (b) $R_{D^{\ast}}$. The black hatched region corresponds to 1$\sigma$ deviation from experimental results and the region enclosed by the uppermost and lowermost black lines corresponds to 2$\sigma$ deviation from experimental results.}
\label{fig:RD} 
\end{figure}
Unlike $b\to s$ transition, the $b\to c$ transition can occur in the SM, albeit CKM suppressed, in tree level, i.e., $n=0$ is allowed in the right panel of Figure~\ref{fig:b2smu}. Thus the nmUED contribution of higher KK mode $W$-bosons in the $b\to c$ process, and hence in $R_{D^{(\ast)}}$, is quite small which is evident from the Figure~\ref{fig:RD} where not much deviation is found in $R_{D^{(\ast)}}$ even after considering BLKT values up to 20. This corroborates similar observations made in the context of warped extra dimensional set-up~\cite{DAmbrosio:2017wis}. In the rest of the paper, we ignore the observables $R_{D^{(\ast)}}.$
 

\section{Dark Matter}
\label{sec:dm}
The preservation of KK parity due to the equal-strength BLKT parameters at the two boundary points implies the stability of the lightest KK particle (LKP). Since in the nmUED set-up the BLKT parameters determine the mass spectrum as well as the couplings, the identity of the LKP can vary depending on the choice of BLKT parameters. In the mUED its always the first KK level photon $A^{(1)}$, which is generally denoted in the literature by $B^{(1)}$ since this is the dominant part in $A^{(1)}$~\cite{Servant:2002aq, Servant:2002hb, Burnell:2005hm, Kong:2005hn}. In nmUED, however, $Z^{(1)}$ (i.e., $W_{3}^{(1)}$), $H^{(1)}$ can also be viable LKPs apart from the usual $A^{(1)}$. The detailed studies of dark matter in the nmUED has been performed in~\cite{Datta:2013nua, Flacke:2017xsv}. In this paper we will stick to the case where $A^{(1)}$ is the DM candidate which is mainly the $B^{(1)}$. Since for a specific value of the compactification radius, the KK masses decrease with increasing values of BLKT parameters, we have to take $r_{g} > r_{f_{i}}$ to avoid fermion LKP. Also in our analysis we do not take into account the co-annihilation processes involving $W_{3}^{(1)}$, $H^{(1)}$ etc. Actually we stick to the conservative overclosure bound only when considering the parameter space, i.e., we demand the relic density of the $A^{(1)}$ to be just less than the observed relic density of the universe.   
The relic density of $A^{(1)}$ can be written, by solving the corresponding Boltzmann equation with appropriate assumptions, as
\begin{equation}
\Omega_{A^{(1)}}h^2 \approx \frac{1.04\times 10^9}{M_{\rm Pl}}\frac{x_F}{\sqrt{g_{\star}(x_F)}}\frac{1}{a+\frac{3b}{x_F}}\;,
\label{eq:relicEqn}
\end{equation}
where $M_{\rm Pl}$ is the Planck mass, $x_{F}(=m_{A^{(1)}}/T_{F})$ represents the freeze-out temperature which we take $\sim$ 25~\cite{Servant:2002aq, Kong:2005hn}, the quantities $a$ and $b$ are the coefficients of the non-relativistic expansion of the thermally averaged annihilation cross section, i.e., $\langle \sigma v \rangle = a + b\langle v^{2} \rangle + \mathcal{O}(\langle v^{4} \rangle) \approx a + 6b/x$ and lastly the quantity $g_{\star}$ is the total number of effective degrees of freedom. 
The main annihilation channels of $A^{(1)}$ are $A^{(1)}A^{(1)} \to f\bar{f}$ and $A^{(1)}A^{(1)} \to HH^{\ast}$, where $f$ represents the SM fermions and $H$ is the SM Higgs. The thermally averaged cross section of these annihilation channels can be expanded in terms of the coefficients $a$ and $b$ which via Eq.~(\ref{eq:relicEqn}) gives the estimation of relic density. In our analysis we also considered the co-annihilation channels with gauge or scalar particles in the initial states, such channels are: (i) $A^{(1)}W_{3}^{(1)}\to f\bar{f}$, (ii) $A^{(1)}W_{3}^{(1)}\to HH^{\ast}$, (iii) $A^{(1)}W^{(1)\pm}\to f\bar{f^{\prime}}$, (iv) $A^{(1)}G^{(1)}\to W^{\mp}G^{\pm}$, (v) $A^{(1)}G^{(1)\pm}\to Z(\gamma)G^{\pm}$. We have used the expressions for cross sections of these processes from~\cite{Kong:2005hn}. 

The DM direct detection searches can put stringent constraints on the parameter space of the model. This can even be more crucial given the ever-increasing precision of modern direct detection experiments like, LUX, Xenon1T etc.~\cite{Akerib:2017kat, Aprile:2018dbl}. The scattering cross section of the DM $A^{(1)}$ off nuclei is ultimately related to its scattering from quarks. In the non-relativistic limit the total cross section can have both the spin-independent and spin-dependent parts. The details of the analysis of the direct detection constraints are discussed in~\cite{Datta:2013nua, Flacke:2017xsv}.
\begin{figure}[h]
	\centering
	 \subfloat[]{
	\includegraphics[scale=0.5]{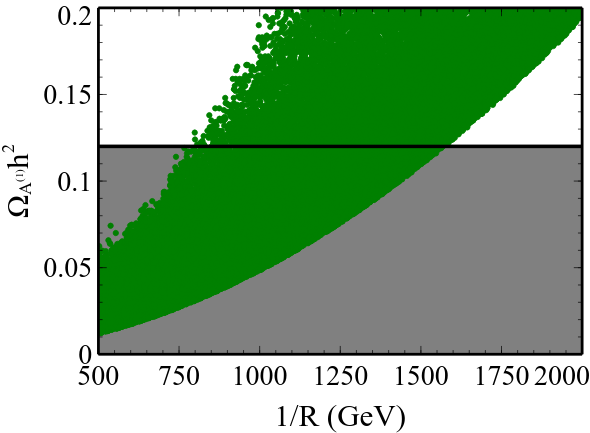}
	}~~~~
	 \subfloat[]{
	\includegraphics[scale=0.5]{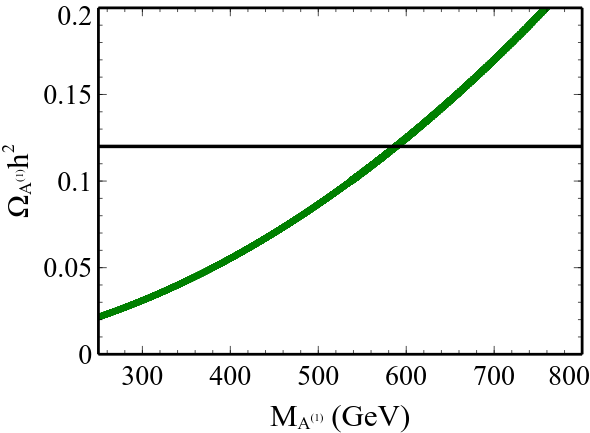}	
	}
	\caption{(a) Allowed parameter points in the relic density, $\Omega_{A^{(1)}}h^2$ vs $1/R$ plane. The black line corresponds to the overclosure bound 0.12~\cite{Ade:2015xua}. (b) The LKP $A^{(1)}$ dependence on the relic density. The overclosure bound fixes the mass of the LKP $M_{A^{(1)}} \sim 600$ GeV.}
	\label{fig:relic}
\end{figure}

We generate BLKT parameters randomly in the range $(0, 8)$ and $1/R$ in the range $(500, 2000)$ GeV and calculate $\Omega_{A^{(1)}}h^2$ using Eq.~\eqref{eq:relicEqn} for each set. In Figure~\ref{fig:relic} we show the allowed points which are also allowed by direct detection constraints from LUX (spin -dependent)~\cite{Akerib:2017kat} and Xenon1T (spin-independent)~\cite{Aprile:2018dbl}. We observe that a large number of parameter points remain allowed by the overclosure bound. However, it sets strict upper limits of 1/R at $\sim 1.6$ TeV and $M_{A^{(1)}}$ at $\sim 600$ GeV.

\section{Phenomenological Constraints}
\label{sec:phenoconstraints}

\subsection{Higgs Data}
\label{sbsec:higgs}
The production of Higgs boson at the LHC dominantly occurs through the gluon fusion process $gg \to H$ which is driven by fermion triangle loops, and within the SM the top quark loop dominates due to its large Yukawa coupling with the Higgs. The subsequent important decay of the SM Higgs to the di-photon is also a loop induced process where SM fermions and $W$-boson run in the triangle loop. In the nmUED, the KK modes of SM fermions as well as $W$-boson can contribute in these loop induced processes and can potentially modify the effective couplings. One can thus put stringent constraints on the model parameter space from the observed Higgs couplings~\cite{Kakuda:2013kba, Dey:2013cqa, Flacke:2013nta, Datta:2013xwa, Bhattacharyya:2009nb, Dey:2015pba}. The cross section and decay width for the Higgs including the KK contributions can be written as,
\begin{align}
\sigma_{gg\to H} &= \frac{G_{F}\alpha_{s}^{2}m_{H}^{3}}
                    {16\sqrt{2}}
                   \left|\frac12 A_{t}(\tau_{t}) + 
                         C(r_{\rm NP})\mathcal{A}_{\rm NP}
                         (\tau_{\rm NP})\right|^{2},\\
\Gamma_{H\to \gamma \gamma} &= \frac{G_{F}\alpha^{2}m_{H}^{3}}
               {128\sqrt{2}}\left|
               A_{W}(\tau_{W}) 
               + 3\left(\frac23\right)^{2}A_{t}(\tau_{t}) 
               + N_{c,\text{NP}}Q_{\rm NP}^{2}
                 \mathcal{A}_{\rm NP}(\tau_{\rm NP}) )\right|^{2},
\end{align}
where $\tau_{j} = m_{H}^{2}/4m_{j}^{2}$, $C(r_{\rm NP})$ is the $SU(3)$ colour factor, $N_{c,\text{NP}}$ is the number of colour states of the new physics (NP) particles, $Q_{\rm NP}$ is the electric charge of the NP particle in the loop. In the light Higgs limit ($m_{H} \ll 2m_{j}$) $A_{t}\sim 4/3$, $A_{W}\sim -7$. The quantity $\mathcal{A}_{\rm NP}$ is defined as,
\begin{align}
\mathcal{A}_{\rm NP}(\tau_{\rm NP}) = v\frac{\partial}{\partial v}
            \text{log}\left[\text{det} 
            \left(\mathcal{M}(v)\right)\right]
            A(\tau_{\rm NP}), 
\end{align}
where $\mathcal{M}(v)$ is the mass matrix for respective particles. Clearly, $\mathcal{A}_{\rm NP}$ contains only contribution from top quark for the $\sigma_{gg\to H}$ but both top quark and $W$ boson for $\Gamma_{H\to \gamma \gamma}$.
%
%

%
With these one can now calculate the dimensionless parameters $c_{gg} = \sigma_{gg\rightarrow h}^{NP}/ \sigma_{gg\rightarrow h}^{SM}$ and $c_{\gamma\gamma} = \Gamma_{gg\rightarrow h}^{NP}/ \Gamma_{gg\rightarrow h}^{SM}$ to compare the modification induced by the KK modes with the experimentally observed values. Using the values obtained in~\cite{Ellis:2013lra}, we have $\sqrt{c_{gg}} = 0.88\pm 0.11$ and $\sqrt{c_{\gamma \gamma}} = 1.18\pm 0.12$. In our analysis we will use this constraint.

\subsection{Oblique Parameters}
\label{sbsec:ewpo}
The oblique corrections to the electroweak gauge boson propagators, incarnated in the Peskin-Takeuchi parameters a.k.a., $S$, $T$ and $U$ parameters, generally put strong constraints on the BSM models. In nmUED these electroweak constraints are discussed in~\cite{Flacke:2008ne, Datta:2013yaa, Datta:2015aka, Flacke:2013pla, Dey:2016cve}. Here we briefly discuss them for completeness. 
In KK parity conserving nmUED the Fermi constant $G_{F}$ receives corrections due to the presence of tree-level coupling between second KK level gauge boson and the SM fermions. This correction can be presented as,
\begin{equation}\label{precision}
\begin{split}
G_F &= G_F^0 + \delta G_F\\
    &= \frac{g^2}{4\sqrt{2}M_W^2} + 
       \sum_{\substack{k\geq 2\\ k\in \text{even}}}
       \frac{\left(g I^{(k)}\right)^2}{4\sqrt{2}M_{W^{(k)}}^2},
\end{split}
\end{equation}
where $G_F^0$ ($\delta G_{F}$) represents the $s$-channel SM (even KK mode) $W^{\pm}$-boson exchange, and $I^{(k)}$ is the relevant overlap integral shown in Eq.~\eqref{eq:ovint}. With these quantities the non-zero Peskin-Takeuchi parameters can be written as~\cite{Datta:2013yaa, Datta:2015aka, Dey:2016cve}, 
\begin{align}
T_{\rm nmUED}=-\frac{1}{\alpha}\frac{\delta G_F}{G_F}, 
\quad
U_{\rm nmUED}=\frac{4\sin\theta_W^2}{\alpha}\frac{\delta G_F}{G_F}.
\end{align}
The most recent fit to the electroweak precision data given by \texttt{Gfitter} group~\cite{Baak:2014ora},
\begin{align}
S = 0.05 \pm 0.11,~
T = 0.09 \pm 0.13,~
U = 0.01 \pm 0.11.
\end{align}
We can now perform the model parameter scan following the prescription of~\cite{Dey:2016cve} by defining the $\chi^{2}$ in terms of the covariance matrix. For a maximal $2\sigma$ ($3\sigma$) deviation, we need $\chi^{2}\leq 6.18$ (9.21) given the two degrees of freedom. 


\subsection{Dilepton bounds}
\label{sbsec:dilep}
The presence of KK number violating (but KK parity conserving) interactions can result in the single production of second KK excitation of gauge bosons. Resonant production of the second level KK gauge boson and its subsequent decay to SM fermions at the LHC, i.e., $pp \to X^{(2)} \to l^{+}l^{-}$ proceeds via couplings  of the form $g_{X^{(2)}f\bar{f}}$. The dilepton searches at the LHC can put stringent constraints on these second level KK gauge bosons. Recent ATLAS~\cite{ATLAS-CONF-2016-045} and CMS~\cite{CMS-PAS-EXO-16-031} searches have been used to put constraint on the parameter space of nmUED in~\cite{Flacke:2017xsv}. Without taking recourse to the extensive collider simulations we simply translate the bounds from~\cite{Flacke:2017xsv} on our simulated sample points. We compare the coupling $g_{X^{(2)}f\bar{f}}$ in our model with the constraint given in~\cite{Flacke:2017xsv} for a given $M_{X^{(2)}}$ to determine the exclusion from LHC dilepton searches. Recasting other collider studies \cite{Datta:2012tv, Datta:2013yaa} of nmUED is beyond the scope of this work. However, we expect enough freedom exists within the framework to evade these constraints due to the presence of non-universal BLKT parameters.

\section{Constraints on the Parameter Space}
\label{sec:results}
In this section we discuss the constraints on the parameter space obtained by imposing various phenomenological constraints  elaborated in the preceding sections.
\begin{figure}[t]
\centering
 \subfloat[]{
  \includegraphics[scale=0.5]{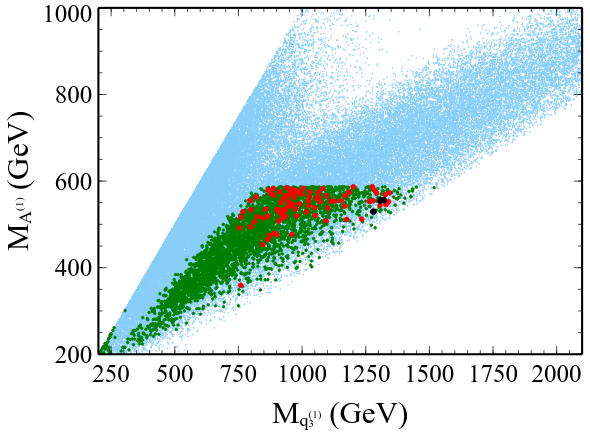}
 } ~~~~
 \subfloat[]{
  \includegraphics[scale=0.5]{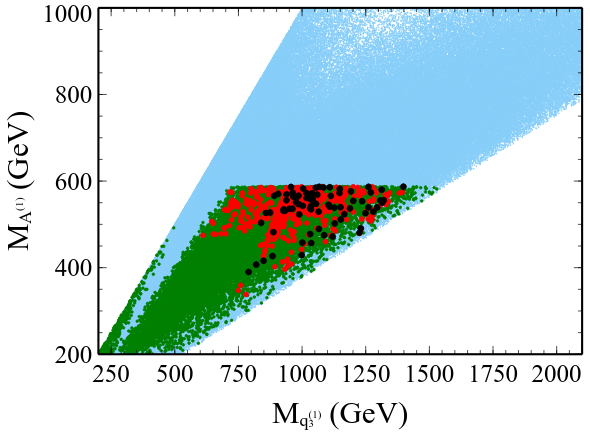}
 }
\caption{Allowed parameter points in the $M_{A^{(1)}}$-$M_{q_{3}^{(1)}}$ plane. The blue points represent the points allowed by (a) 1$\sigma$ and (b) 2$\sigma$ values of $R_{K}$ and $R_{K^{\ast}}/R_{K}$. The green points show the allowed points after additionally imposing the DM constraints i.e., the overclosure bound and direct detection cross sections. The red points are allowed after additionally imposing the Higgs decay width constraints. Finally the black points represent the allowed points after additionally imposing the electroweak precision constraints.}
\label{fig:ma1vsmq31} 
\end{figure} 
We perform a simulation of two million randomly generated parameter points with BLKT parameters in the range $(0, 8)$ and $1/R$ in the range $(500, 2500)$ GeV.  In Table~\ref{tab:bltval}, the Set-I represents the full set of randomly generated parameter points. For each set of input parameters all the  observables are calculated. The various constraints are sequentially imposed on this generated sample points. The fractional cutflow for surviving points  after implementation of subsequent  constraints is depicted in Table~\ref{tab:constraints}. We present the points allowed by $R_K$ , $R_{K^{\ast}}/R_K$, dark matter, Higgs and EWPT constraints in the $M_{A^{(1)}}$-$M_{q_3^{(1)}}$ plane in Figure~\ref{fig:ma1vsmq31}.  We are left with just three points from our sample set that are within $2\sigma$ of $R_{K}$ and $R_{K^{\ast}}/R_K$, and survive the additional imposition of dilepton constraints.  These are shown as benchmark points in Table~\ref{tab:BP}. No point survives all constraints that is simultaneously within the 1$\sigma$ band of $R_K$ and $R_{K^{\ast}}/R_K$. This has to be contrasted with Figure~\ref{sfb:RKvsmKK} where we obtain a large number of points which are within the 1$\sigma$ band of the two flavor observables. This clearly espouses the tension between the flavor observables and precision observables, Higgs data and dark matter observations. We conclude from this generic scan that the surviving points are tuned and accord a marginal improvement in $R_{K^{(\ast)}}.$ 
\begin{table}[t]
\centering
\begin{tabular}{|c|c|c|c|c|c|c|}
\hline
        & \begin{tabular}[c]{@{}c@{}}No. of \\ points (in million)\end{tabular} & $R_{f_{1}}$  & $R_{f_{2}}$  & $R_{f_{3}}$    & $R_{g}$        & \begin{tabular}[c]{@{}c@{}}$1/R$\\ (GeV)\end{tabular} \\ \hline \hline
Set-I  & 2.0                                                & $(0, 8)$     & $(0, 8)$   & $(0, 8)$      & $(0, 8)$      & $(500, 2500)$                                         \\ \hline \hline
Set-II & 0.28                                                        & $(5.5, 6.6)$ & $(4.2, 5.5)$ & $(0.0, 1.0)$ & $(7.3, 8.5)$ & $(1300, 1500)$                                        \\ \hline
\end{tabular}
\caption{Details of the choice of parameter range for numerical scans. Set-II represents a refined scan around BP-III defined in Table~\ref{tab:BP}.}
\label{tab:bltval}
\end{table}
\begin{figure}[t]
\centering
  \includegraphics[scale=0.6]{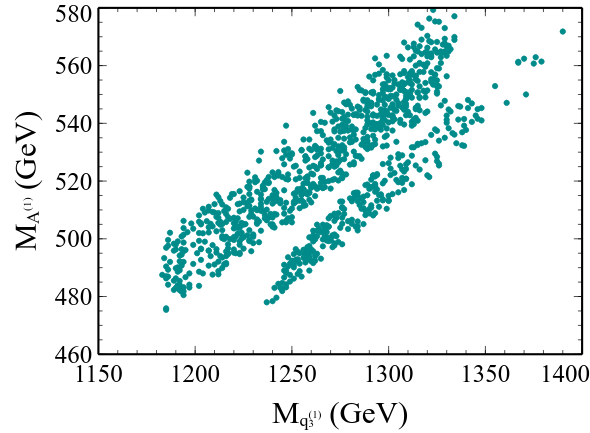}
\caption[]{Allowed parameter points which are within 2$\sigma$ of $R_{K^{(\ast)}}$ and are allowed by other constraints are presented in the parameter space of $M_{A^{(1)}}-M_{q_3^{(1)}}.$} 
\label{fig:all} 
\end{figure}  
We identify  BP-III of Table~\ref{tab:BP} as a point relatively insulated from collider constraint and perform a finer scan around this point defined by Set-II of Table~\ref{tab:bltval}. We show the allowed points in the parameter space of $M_{A^{(1)}}-M_{q_3^{(1)}}$ in Figure~\ref{fig:all}. These points represent a section of the parameter space that remains allowed by all constraints and can explain the $R_{K^{(\ast)}}$ observation within $2\sigma$.\\\\\\

A qualitative analysis of the cuts imposed points towards the region of the parameter space that has survived. The arguments are as follows: 
\begin{enumerate}
\item The tree level flavor violation in the quark sector (Fig.~\ref{sfa:b2smu}) is proportional to $|R_{f_2}-R_{f_3}|.$ Thus a sizeable flavor violation indicated by $R_{K^{\ast}}$ prefers a considerable difference between the values of $R_{f_2}$ and $R_{f_3}$.
\item In Eqs.~(\ref{eq:wilsoncoeff}) the vectorial Wilson coefficient $|\Delta C_{9}|$ dominates over $|\Delta C_{10}|$. From Eq.~(\ref{rk}) it is evident that for $R_K < 1$, one needs $|\Delta C_9^{\mu}|$  to be greater than $|\Delta C_9^e|.$ And since $|\Delta C_9^i|$ is inversely proportional to $R_{f_i}$, we require $R_{f_1}$ to be greater than $R_{f_2}.$
\item The increasingly SM like Higgs data tunes $R_{f_3}$ to small positive values.
\item The allowed parameter space from precision data closes down the difference between $R_{f_2}$ and $R_{\phi}$ as evident from Eqs.~(\ref{precision}) and (\ref{eq:ovint}). 
\end{enumerate}
As we have assumed $R_{\phi}$ to be greater than $R_{f_i}$ to get a bosonic LKP, following the above arguments we conclude the following hierarchy structure of the BLKT parameters: $R_{\phi}\geq R_{f_1} \geq R_{f_2} > R_{f_3}$, which is exactly the distribution we have found in our benchmark points.

%
\begin{table}[t]
\centering
\begin{tabular}{|c|c|c|c|c|c|}
\hline
Total                                                             & $R_{K}^{2\sigma}$,$(R_{K^{\star}}/R_K)^{2\sigma}$ & DM           & Higgs          & EWPO                  & Dilepton               \\ \hline
1 & 0.2          & 0.01       & $2\times 10^{-4}$        & $4\times 10^{-5}$ & $2\times 10^{-6}$ \\ \hline \hline
Total                                                             & $R_{K}^{1\sigma}$,$(R_{K^{\star}}/R_K)^{1\sigma}$ & DM           & Higgs          & EWPO                  & Dilepton               \\ \hline
1                                                             & 0.03           & $1\times 10^{-3}$       & $4\times 10^{-5}$       & $2\times 10^{-6}$ & 0                    \\ \hline \hline
\end{tabular}
\caption{Survival fraction after subsequent imposition of various constraints.
	}
\label{tab:constraints}
\end{table}

\begin{table}[t]
\centering
\begin{tabular}{|c|c|c|c|c|c|c|c|c|c|}
\hline
       & $R_{g}$ & $R_{f_{1}}$ & $R_{f_{2}}$ & $R_{f_{3}}$ & \begin{tabular}[c]{@{}c@{}}$M_{A^{(1)}}$\\ (GeV)\end{tabular} & \begin{tabular}[c]{@{}c@{}}$M_{q_{3}^{(1)}}$\\ (GeV)\end{tabular} & $R_{K}$ & $R_{K^{\ast}}/R_{K}$ & $\Omega h^{2}$ \\ \hline \hline
BP-I   & 5.098    & 3.922        & 2.966        & 0.2385     & 540.8                                                           & 1108                                                              & 0.924   & 1.034                & 0.101            \\ \hline
BP-II  & 6.404   & 4.997       & 4.087        & 0.0281     & 539.1                                                           & 1295                                                              & 0.9358     & 1.028              & 0.101            \\ \hline
BP-III & 7.788   & 6.102       & 4.746        & 0.3496     & 525.8                                                           & 1249                                                              & 0.9279     & 1.032              & 0.0958            \\ \hline
\end{tabular}
\caption{The three benchmark points that are allowed by all the constraints.}
\label{tab:BP}
\end{table}

\section{Summary and Conclusion}
\label{sec:concl}
Non-minimal version of the UED has a rich phenomenology owing to the presence of boundary localized parameters which play a consequential role in determining not only the mass spectrum of the KK particles but also significant deviations in the relevant couplings. In this paper we present a variant of nmUED with non-universal BLKT parameters with  an $U(2)_{Q_{L}} \otimes U(2)_{u_{R}} \otimes U(2)_{d_{R}}$ symmetry in the  fermionic sector. 
This  flavour maximal set-up implies new flavour violating contributions mediated by the KK gauge bosons while contributions from the  Yukawa sector remain suppressed by the Yukawa couplings. Utilizing this flavour violation we explore  the possibility of reconciling  the recent discrepancy between the observation and SM expectation in flavour observables $R_{K^{(\ast)}}$ and $R_{D^{(\ast)}}$ while conforming to other collider and cosmological constraints. We perform an extensive scan of the parameter space to find the allowed regions that is in consonance with all constraints. We present an updated study of the dark matter aspects of the model by using the relic density observations from Planck and latest direct detection cross section results from LUX and Xenon1T which already constrains the $B^{(1)}$ LKP mass in the ballpark of 600 GeV. The other observables which we consider include (i) Higgs data, namely the $H\to \gamma \gamma$ and $gg\to H$ channels where the effect of KK modes will be most dominant; (ii) electroweak precision observables a.k.a. oblique parameters; (iii) bounds from dilepton searches of ATLAS and CMS.  
We find that while the new contributions are significant in bridging the gap for the $R_{K^{(\ast)}}$, the improvements in terms of $R_{D^{(\ast)}}$ is numerically insignificant. The extensive scan reveals that even with all the flexibility afforded by non-universality of the BLKT parameters only a fine-tuned region of parameter space is capable of evading the phenomenological constraints while remaining within $2 \sigma$ of the measured values of $R_K$ and $R_{K^{\ast}}/R_K.$  While we find no parameter point that is within $1 \sigma.$ 
As non-observation of new physics at the LHC pushes the KK scale ever higher it is in imminent danger of being severely constrained from the upper bound on the LKP arising from the dark matter relic density over-closure bound. It is evident from the analysis presented in this paper that the requirements to match the flavour discrepancies are in tension with constraints from precision physics, Higgs data and dark matter observations. Relaxation of this tension requires further generalisation of the BLKT framework. For example, one might consider different BLKT parameter for chiral components of the same flavour or explicit violation of KK parity by introducing different BLKT parameter at different branes. A detailed and rigorous study of the collider, flavour and cosmological aspects of such a generalised framework would be in order.

\paragraph*{Acknowledgements\,:}
We acknowledge Gudrun Hiller for useful correspondence. SD acknowledges MHRD, Govt. of India for the research fellowship. UKD acknowledges the support from Department of Science and Technology, Government of India under the fellowship reference number PDF/2016/001087 (SERB National Post-Doctoral Fellowship). TSR is partially supported by the Department of Science and Technology, Government of India, under the Grant Agreement number IFA13-PH-74 (INSPIRE Faculty Award).

\appendix

\section{Model Details}
\label{appn:model}
The five-dimensional four component quark fields ($Q, U, D$) can be represented as their corresponding Kaluza-Klein (KK) excitations as,
\begin{subequations}
\begin{align}
Q(x,y) &=  N_{Q}^{(0)}\,Q_{L}^{(0)}+ \sum^{\infty}_{n=1}\left[ Q_{L}^{(n)}(x) f_{L}^{(n)}(y) + Q_{R}^{(n)}(x) g_{L}^{(n)}(y) \right], \\
U(x,y) &= N_{Q}^{(0)}\,U_{R}^{(0)}+ \sum^{\infty}_{n=1}\left[ U_{L}^{(n)}(x) f_{R}^{(n)}(y) + U_{R}^{(n)}(x) g_{R}^{(n)}(y)\right], \\
D(x,y) &= N_{Q}^{(0)}\,D_{R}^{(0)}+ \sum^{\infty}_{n=1}\left[ D_{L}^{(n)}(x) f_{R}^{(n)}(y) + D_{R}^{(n)}(x) g_{R}^{(n)}(y)\right].
\end{align} 
\end{subequations}
In the effective 4D theory the zero modes of $Q$ generates the $SU(2)_{L}$ doublet quarks and the zero modes of $U$ ($D$) are identified with the up (down) type singlet quarks, i.e. after compactification and orbifolding the zero modes of $Q$ gives the left-handed doublet comprising of SM $t_{L}$ and $b_{L}$, whereas $t_{R}$ and $b_{R}$ emerges from the $U$ and $D$ respectively. The compact form of quark doublet is $Q \equiv (Q_{i}, Q_{j})^{T}$, where $i$ and $j$ correspond to up type and down type quark respectively. $N_{Q}^{(0)}$ is the normalization constant of the fermionic wave functions for zero-mode. In Eq.~(\ref{eq:quarkAction}) the terms containing the parameter $r_{f}$ are the boundary localised kinetic terms (BLKTs). In the mUED, $r_{f}$ is assumed to be vanishing, thus by setting the BLKT parameters to zero one can translate from nmUED to mUED. Also note that the Latin indices in Eq.~(\ref{eq:quarkAction}) run from 0 to 4 whereas the Greek indices from 0 to 3. We will be using the  metric convention $g_{MN}\equiv \rm{diag}(+1,-1,-1,-1,-1)$. The covariant derivative, $D_M\equiv\partial_M-i\widetilde{g}W_M^a T^a-i \widetilde{g}^\prime B_M Y$, where $\widetilde{g}$ and $\widetilde{g}^\prime$ are the 5D gauge coupling constants of $SU(2)_L$ and $U(1)_Y$, respectively, and $T^a$ and $Y$ are the corresponding generators. The five-dimensional gamma matrices are $\Gamma^{M}=(\gamma^{\mu},-i\gamma_{5})$.
The variation of the action with appropriate boundary conditions lead to the $y$-dependent mode functions $f$ and $g$ as,
\begin{eqnarray}
\label{eq:modefunctions}
f_{L}(y) = g_{R}(y) = N_{Q}^{(n)} \left\{ \begin{array}{rl}
                \displaystyle \frac{\cos[M_{Q_{n}} \left (y - \frac{\pi R}{2}\right)]}{C_{Q_{n}}}  &\mbox{$\forall$ $n$ even,}\\
                \displaystyle \frac{{-}\sin[M_{Q_{n}} \left (y - \frac{\pi R}{2}\right)]}{S_{Q_{n}}} &\mbox{$\forall$ $n$ odd,}
                \end{array} \right. 
\end{eqnarray}
and
\begin{eqnarray}
\label{eq:modefunctionsodd}
g_{L}(y) = f_{R}(y) = N_{Q}^{(n)} \left\{ \begin{array}{rl}
                \displaystyle \frac{\sin[M_{Q_{n}} \left (y - \frac{\pi R}{2}\right)]}{C_{Q_{n}}}  &\mbox{$\forall$ $n$ even,}\\
                \displaystyle \frac{\cos[M_{Q_{n}} \left (y - \frac{\pi R}{2}\right)]}{S_{Q_{n}}} &\mbox{$\forall$ $n$ odd,}
                \end{array} \right. 
\end{eqnarray}
with
\begin{equation}
C_{Q_{n}} = \cos\left(\frac{M_{Q_{n}} \pi R}{2}\right),\, 
\,\,\,\,
S_{Q_{n}} = \sin\left(\frac{M_{Q_{n}} \pi R}{2}\right).
\end{equation}
The mode functions $f$ and $g$ satisfy the orthonormality conditions,
\begin{equation}\label{eq:orthonorm}
\int dy \left[1 + r_{f}\{ \delta(y) + \delta(y - \pi R)\}
\right] ~k^{(m)}(y) ~k^{(n)}(y) = \delta^{mn} = \int dy ~l^{(m)}(y) ~l^{(n)}(y)
\end{equation}
where, $k$ can be $f_{L}$ or $g_{R}$ and $l$ corresponds to $g_{L}$ or $f_{R}$. From the above condition one can obtain the normalization factors as
\begin{equation} \label{eq:norm}
N_{Qn} = \sqrt{\frac{2}{\pi R}}\left[ \frac{1}{\sqrt{1 + \frac{r_f^2 M_{Qn}^{2}}{4} 
+ \frac{r_f}{\pi R}}}\right].
\end{equation}
For zero-mode, the normalization constant is given by
\begin{equation} \label{eq_norm_zero}
N_{Q0} = \frac{1}{\sqrt{r_{f}+\pi R}}.
\end{equation}
Note that $r_{f} = 0$ implies the usual (m)UED normalization $\sqrt{2/(\pi R)}$ for $n$-th mode, whereas for zero-mode it is $\sqrt{1/(\pi R)}$. The $y$-profile solutions for the gauge field $V^{\mu}$ (with $V= W, Z, \gamma$) is given by the Eqs.~(\ref{eq:modefunctions}) with $M_{Q_{n}}$ being replaced by $M_{\Phi_{n}}$. The $M_{\Phi_{n}}$ is the solution of transcendental equations (Eq.~(\ref{eq:transc})) for gauge fields with BLKT parameter $r_{\phi}$ instead of $r_{f}$. The orthonormality condition will be similar to that of $k^{(i)}(y)$ of Eq.~(\ref{eq:orthonorm}). The corresponding normalization factor is given by $N_{\Phi 0}= 1/\sqrt{r_{\phi}+\pi R}$.

\section{Flavour Violation: Gauge and Yukawa Sector}
\label{appn:flav}
Here we give a brief outline of the flavour violation that can arise in the model for different choices of model parameters. 
In the gauge sector the relevant interactions of $W^{(n)\pm}$ and
$Z^{(n)}$ are shown in Eqs.~\eqref{eq:LEW} and \eqref{eq:LW}. 
After EWSB, the quark mass matrices are diagonalised by unitary 
matrices $V_{u}$ and $V_{d}$ which are the constituents of the CKM 
matrix $V_{\rm CKM} = V_{u}^{\dagger}V_{d}$. Now, flavour 
violation occurs both in the neutral and charged currents in the 
mass eigenstate by $\left(V_{d_{L,R}}^{\dagger} \mathcal{I}^{(n)} 
V_{d_{L,R}}\right)$ and $V_{u_{L}}^{\dagger}\mathcal{I}^{(n)}V_{d_{L}}$ respectively, where $\mathcal{I}^{(n)} = \text{diag}\{I^{(n)}(R_{g},R_{f_{1}}), I^{(n)}(R_{g},R_{f_{2}}), I^{(n)}(R_{g},R_{f_{3}})\}$. Evidently, the flavour violation is generated due to the fact that the matrix $\mathcal{I}^{(n)}$ is diagonal but not proportional to identity matrix. This is mandated by the generation-wise different BLT parameter, i.e., $r_{f_{1}} \neq r_{f_{2}} \neq r_{f_{3}}$.

The Yukawa Lagrangian for the charged scalar part in the presence of the boundary localised Yukawa term can be written as,
\begin{align}
\label{eq:yukchg}
\mathcal{L}_{\rm Yuk}^{\rm charged} = \int d^{4}x \int_{0}^{\pi R}
                        dy \left[1 + r_{y}\{\delta(y) 
                        + \delta(y-\pi R)\} \right]
                        \left(
                          - \tilde{y}_{d} \bar{Q}_{L} 
                                          \phi^{(n)+} D
                        \right)\;.
\end{align}
Here we write the interactions for the down sector; the discussion for the up sector follows similarly. Also, we do not take any flavour structure in $r_{y}$ itself. The 5D Yukawa $\tilde{y}_{d}$ and its 4D counterpart are related by the relation,
\begin{align}
\tilde{y}_{d_{i}} = \frac{y_{d_{i}}}{\pi R} 
                    \left(r_{f_{i}} + \pi R\right)
                    \sqrt{r_{\phi} + \pi R}\;,
\end{align}
which can be obtained by taking $r_{y} = 0$ and considering the zero mode wave functions of corresponding fields with appropriate normalisation factors. Now, to check the flavour violation in the Yukawa sector, we consider the coupling between the higher KK mode charged scalar and the zero mode fermions. This coupling can be obtained from Eq.~\eqref{eq:yukchg}. 
By taking appropriate normalisations into account one can show this coupling to be,
\begin{align}
\label{App:B2}
-y_{d_{i}} \frac{\sqrt{r_{\phi} + \pi R}}{\pi R}
                \int_{0}^{\pi R} dy 
                        f_{\Phi}^{(2)}(y)\;.
\end{align}
Since this coupling is $r_{f_{i}}$ independent there will be no flavour violation.

 We can also take $r_{f} \neq 0 \neq r_{y}$, where $r_{y}$ is same for all three generations. In that case, the 5D Yukawa $\tilde{y}_{d}$ coupling is related to its 4D counterpart by,
\begin{align}
\label{App:B3}
\tilde{y}_{d_{i}} = y_{d_{i}}
                   \frac{ \left(r_{f_{i}} + \pi R\right)}{\left(r_{y} +\pi R\right)}
                    \sqrt{r_{\phi} + \pi R}\;.
\end{align}
For the above condition, the Eq.~(\ref{App:B2}) can be rewritten as
\begin{align}
\label{App:B4}
-y_{d_{i}} \frac{\sqrt{r_{\phi} + \pi R}}{r_{y} + \pi R}
                \int_{0}^{\pi R} dy \left[1 + r_{y}\{\delta(y) 
                        + \delta(y-\pi R)\} \right]
                        f_{\Phi}^{(2)}(y)\;
\end{align}
which is again independent of $r_{f}$, so there is no flavour violation. But if we take $r_{y}$ different for all generations or set $r_{y_{i}} = r_{f_{i}}$, then we will get flavour violations in the Yukawa sector due to the presence of the terms like $1/\left(r_{{y}_{i}} +\pi R\right)$ or $1/\left(r_{{f}_{i}} +\pi R\right)$ respectively in the $\mathcal{I}^{(n)}$ matrix. In this article, we actually consider the second one, i.e. $r_{f} \neq 0 \neq r_{y}$ and $r_{y}$ being same for all generations, we do not have any flavour violation in the scalar sector.
In passing we mention that the for the BLKT parameter choice, $r_{y} \neq r_{f_{i}}$ there exists KK level mixing in the Yukawa interactions. Due to the KK parity conservation this mixing occurs between even-even modes or odd-odd modes. The mixing angle between the $n$-th and $(n + 2k)$-th ($k\in \mathbb{Z}$) KK level can be estimated as $\theta_{\rm mix}\sim I(r_{y},r_{f_{i}},n,k) m_{f_{i}}R/2k$, where $I(r_{y},r_{f_{i}},n,k)$ is the corresponding overlap integral. This mixing angle is suppressed by the new physics scale, see~\cite{Datta:2012tv, Dey:2013cqa} for details.



\bibliographystyle{JHEP}
\bibliography{ref.bib}

\end{document}